\documentclass[aps,prd,preprint,superscriptaddress,nofootinbib,showpacs]{revtex4-2} 
\usepackage{graphicx,amsmath,amsfonts,amssymb,revsymb,dcolumn,relsize,hyperref}
\hypersetup{colorlinks=true,citecolor=blue,urlcolor=blue,linkcolor=blue}

\newcommand{\be}{\begin{equation}}
\newcommand{\ee}{\end{equation}}
\newcommand{\bea}{\begin{eqnarray}}
\newcommand{\eea}{\end{eqnarray}}

\begin{document}

\title{Free fall in modified symmetric teleparallel gravity}

\author{Nivaldo A. \surname{Lemos}} \email{nivaldolemos@id.uff.br}
\affiliation{Instituto de F\'{\i}sica, Universidade Federal Fluminense,
Av. Litor\^anea, S/N \\
24210-340 Niter\'oi -- RJ, Brazil }




\renewcommand{\baselinestretch}{0.96}

\date{\today}

\begin{abstract}
The status of the equivalence principle in modified symmetric teleparallel gravity is examined. In this theory, minimum length  geodesics are distinct from autoparallel geodesics, that is, the ``shortest'' paths are not the ``straightest'' paths.  We show that a standard argument that singles out metric geodesics in general relativity does not apply in  modified symmetric teleparallel gravity. This is because the latter theory does not obey the equivalence principle in the sense of Weinberg. We argue, however,  that the structure of the theory makes it inevitable that a freely falling test particle follows a shortest path, a geodesic of the metric.  The geodesic equation that governs the motion of a freely falling test particle involves the Levi-Civita connection, not some other connection obtained by solving the connection field equations of the theory. This also has bearing  on whether, under appropriate conditions,  modified symmetric teleparallel gravity is fully  equivalent to general relativity.

\vspace{1cm}

\end{abstract}




\maketitle
\newpage

\section{Introduction}

In general relativity the equivalence principle entails that a test particle that is not subject to any nongravitational forces follows a geodesic of spacetime. In Einstein's theory of gravity geodesics  are equivalently defined either as the shortest paths or the autoparalell paths. Modified symmetric teleparallel gravity, also known as $f(Q)$ gravity, is a new theory of gravity  formulated in terms of the nonmetricity tensor \cite{nester}, which measures the departure from zero of the covariant derivative of the metric tensor (see \cite{Lavinia-review} for a recent  extensive review). A nonvanishing nonmetricity implies that  the connection that characterizes parallel transport is not the Levi-Civita connection (Christoffel symbols). The curvature tensor and the torsion tensor  are both postulated to be zero and gravitational effects are encoded in the nonmetricity. The dynamics of modified symmetric teleparallel gravity arises from a modified Einstein-Hilbert action in which the  Ricci scalar  $R$ is replaced by $f(Q)$, where $f$ is a infinitely differentiable function of the nonmetricity scalar $Q$ constructed from the nonmetricity tensor. This modified Einstein-Hilbert action leads to new field equations that are not equivalent to Einstein's equations as long as the second derivative of $f$ does not vanish and $Q$ is not constant. Beside the metric field equations there are separate connection field equations that  also must be satisfied.

Because the nonmetricity tensor does not vanish, in modified symmetric teleparallel gravity there are two inequivalent classes of geodesics, namely the shortest paths and the autoparallel or straightest paths. This raises the question whether in modified symmetric teleparallel gravity a freely falling test particle follows a shortest or a straightest path.

In Section \ref{Essentials}   we present a very concise exposition of the essential aspects of  modified symmetric teleparallel gravity. In Section \ref{two-kinds-geodesics} we describe the two different classes of geodesics. In Section \ref{freefall} we examine whether by a standard argument, which is valid in general relativity and picks out the metric geodesics,   one can discriminate between the two candidates to describe the paths of  freely falling test particles. It turns out that the said argument does not apply in modified symmetric teleparallel gravity because the theory violates the equivalence principle in the sense of Weinberg. However,  in Section \ref{Bianchi-identity} we argue that the structure of the theory  singles out the geodesics defined as the shortest paths. Section \ref{Conclusion} is devoted to a few final comments.

\section{Essentials of $f(Q)$ gravity}\label{f(Q)gravity}\label{Essentials}

Modified symmetric teleparallel gravity is the nonlinear extension of the theory of coincident general relativity
 \cite{nester,lavinia1},   also known as symmetric teleparallel equivalent of general relativity. Beside the metric, its other basic object is the nonmetricity tensor $Q_{\alpha \mu \nu}$ which is defined by
\begin{equation}
\label{Nonmetricity}
Q_{\alpha \mu \nu}=\widetilde{\nabla} _\alpha g_{\mu\nu} = \partial_{\alpha}g_{\mu\nu} - \widetilde{\Gamma}^{\lambda}_{\,\,\, \alpha\mu} g_{\lambda\nu}-  \widetilde{\Gamma}^{\lambda}_{\,\,\, \alpha\nu} g_{\mu\lambda}
\end{equation}
and measures the variation of the norm of a vector field under parallel transport specified by the symmetric connection $\widetilde{\Gamma}^{\lambda}_{\,\,\, \mu\nu}$. We reserve the notation $\Gamma^{\lambda}_{\,\,\, \mu\nu}$ for the Levi-Civita connection, whereas $\widetilde{\Gamma}^{\lambda}_{\,\,\, \mu\nu}$ will denote any other  connection used to define parallel transport in the theory. The connection $\widetilde{\Gamma}^{\lambda}_{\,\,\, \mu\nu}$ is not given in advance, it is dynamically determined. The assumption that $Q_{\alpha\mu\nu}$ does not vanish makes room for a different geometric interpretation  of gravitational interactions, focusing on the distortion of the scalar product of vectors \cite{Lavinia-review}. Since  the curvature tensor and the  torsion tensor are assumed to be zero, the connection $\widetilde{ \Gamma}^{\lambda}_{\,\,\, \mu\nu}$ is symmetric and gravitational effects are embodied in the nonmetricity tensor.

The usual covariant derivative associated with the Levi-Civita connection
\begin{equation}
\label{christoffel-symbols}
\Gamma^{\lambda}_{\,\,\,\mu\nu} = \frac{1}{2} g^{\lambda\sigma} \left( \partial_{\mu} g_{\nu\sigma} + \partial_{\nu} g_{\mu\sigma} - \partial_{\sigma} g_{\mu\nu} \right)
\end{equation}
is defined by
\begin{equation}
\label{christoffel-symbols-covariant-derivative}
\nabla_\mu V^{\nu} = \partial_{\mu}V^{\nu} + \Gamma^{\nu}_{\,\,\, \mu\lambda}V^{\lambda}, \qquad \nabla_\mu \omega_{\nu} = \partial_{\mu}\omega_{\nu} - \Gamma^{\lambda}_{\,\,\, \mu\nu}\omega_{\lambda}. 
\end{equation}

For  future reference, let us record the well-known transformation law of any connection under the coordinate transformation $x^{\mu} \to {\hat x}^{\mu} (x)$:
\begin{equation}
\label{connection-transformation}
{\hat \Gamma}^{\lambda}_{\,\,\,\mu\nu}({\hat x}) = \frac{\partial x^{\alpha}}{\partial{\hat x}^{\mu}} \frac{\partial x^{\beta}}{\partial{\hat x}^{\nu}}
\frac{\partial {\hat x}^{\lambda}}{\partial x^{\sigma}}\Gamma^{\sigma}_{\,\,\,\alpha\beta}(x) + \frac{\partial {\hat x}^{\lambda}}{\partial x^{\sigma}}\frac{\partial^2 x^{\sigma}}{\partial {\hat x}^{\mu} \partial {\hat x}^{\nu}}.
\end{equation}
It is worth emphasizing that this holds not only for the Levi-Civita connection but also for any connection $\widetilde{\Gamma}^{\lambda}_{\,\,\,\mu\nu}$ whatsoever.

\subsection{Field equations}

The dynamics of modified symmetric teleparallel gravity stems from the action \cite{lavinia1} 
\begin{equation}
\label{action}
S = \frac{1}{2}\int d^4 x \sqrt{-g}f(Q) + S_m,  
\end{equation}
where $8\pi G/c^4 =1$ and $S_m$ is the matter action.
Setting the variation of this action with respect to the metric equal to zero leads to the metric field equations \cite{cosmology1}
\begin{equation}
\label{f(Q)-equations}
 \frac{2}{\sqrt{-g}}\widetilde{\nabla}_\alpha\bigl(\sqrt{-g}f_QP^{\alpha}_{\,\,\,\,\mu \nu}\bigr)- \frac{1}{2}fg_{\mu \nu}+f_Q\bigl(P_{(\mu \vert\alpha \beta}Q_{\nu )}^{\,\,\,\alpha \beta}-2P_{\alpha \beta (\mu}Q^{\alpha \beta}_{\,\,\,\,\,\,\,\, \nu )}\bigr) = T_{\mu \nu}, 
\end{equation}
where the parentheses surrounding the indices denote symmetrization and the subscript $Q$ means derivative: $f_Q(Q) = f^{\prime}(Q)$. Equating to zero the variation of the action \eqref{action} with respect to the connection, under the constraints that both curvature and torsion vanish,  yields the connection field equations \cite{cosmology1}
\begin{equation}
\label{connection-equations}
\widetilde{\nabla}_{\mu}\widetilde{\nabla}_{\nu} \bigl( \sqrt{-g}f_Q P^{\mu\nu}_{\quad \alpha}\bigr) =0,
\end{equation}
which are dynamical equations for the connection.  To be more precise, the enforcement of the zero curvature and torsion  constraints by means of  Lagrange multipliers  brings about the connection field equations \eqref{connection-equations} and  ensures a manifestly covariant formulation of the theory \cite{Lavinia-review,Lavinia-Palatini}.  It turns out that the theory can also be made covariant,  without changing its physical content, by means of a sort of  St\"uckelberg procedure that involves introducing a set of four auxiliary scalar  fields $\xi^{\mu}$, which in fact represent a possible choice of coordinates \cite{Blixt}.

In the previous equations there appears the nonmetricity conjugate   tensor $P^{\alpha\mu \nu}$ defined as
\begin{equation}
\label{P-alpha-mu-nu}
    P^{\alpha\mu \nu}=-\frac{1}{4}Q^{\alpha\mu \nu}+\frac{1}{2}Q^{(\mu\nu)\alpha}+\frac{1}{4}(Q^{\alpha}-{\tilde Q}^{\alpha})g^{\mu \nu}-\frac{1}{4}g^{\alpha (\mu}Q^{\nu)}.
\end{equation}
The two independent traces of the nonmetricity tensor that appear in the preceding equation are  given by
\begin{equation}
\label{Q-tildeQ}
    Q_{\alpha} = g^{\sigma \lambda}Q_{\alpha \sigma \lambda}, \qquad {\tilde Q}_{\alpha} = g^{\sigma \lambda}Q_{\sigma \alpha \lambda},
\end{equation}
while the   nonmetricity scalar $Q$ is 
\begin{equation}
\label{nonmetricity-scalar-definition}
    Q=-Q_{\alpha \mu \nu}P^{\alpha \mu \nu}.
\end{equation}

\subsection{Equivalence with general relativity}

An interchangeable and more transparent form of the metric field equations \eqref{f(Q)-equations} is \cite{Lin-2021,Zhao-2022}
\begin{equation}
\label{f(Q)-equations-coordinate-basis}
 f_Q {\overset{\mbox{\tiny o}}{G}}_{\mu \nu} + \frac{1}{2}g_{\mu \nu} \bigl( Qf_Q - f)+ 2f_{QQ}\bigl(\partial_{\lambda}Q\bigr) P^{\lambda}_{\,\,\,\, \mu \nu}  = T_{\mu \nu}
\end{equation}
where ${\overset{\mbox{\tiny o}}{G}}_{\mu \nu}$ is the Einstein tensor associated with the Levi-Civita connection.
In this form, it becomes clear that if $f_{QQ}=0$ then $f(Q)= AQ+B$ and at the level of the field equations the theory is equivalent to general relativity (if $B = 0$) or general relativity with a cosmological constant (if $B\neq 0$). Besides, if the nonmetricity scalar $Q$ is constant the metric field equations \eqref{f(Q)-equations-coordinate-basis} are also equivalent to those of general relativity with a cosmological constant, but now for any function $f(Q)$. By way of example, for G\"odel-type metrics in the dynamically admissible coincident gauge defined by $ \widetilde{\Gamma}^{\lambda}_{\,\,\, \mu\nu} =0$ the nonmetricity scalar  $Q$ is constant \cite{Alisson}.


\section{Shortest and straightest paths}\label{two-kinds-geodesics}

The shortest paths arise from the variational principle
\begin{equation}
\label{variational-principle}
\delta \int ds = \delta \int \sqrt{g_{\mu\nu}{\dot x}^{\mu}  {\dot x}^{\nu}} d\tau = 0,
\end{equation}
where 
\begin{equation}
\label{proper-time}
 d\tau^2 = g_{\mu\nu}dx^{\mu}  dx^{\nu} 
\end{equation}
is the proper time along the path. As is well known, this variational principle leads to
\begin{equation}
\label{geodesics-shortest-path}
\frac{d^2x^{\mu}}{d\tau^2} +   \Gamma^{\mu}_{\,\,\, \alpha\beta} \frac{dx^{\alpha}}{d\tau} \frac{dx^{\beta}}{d\tau} =0, 
\end{equation}
where $\Gamma^{\mu}_{\,\,\, \alpha\beta}$ is the Levi-Civita connection. Curves satisfying Eq. \eqref{geodesics-shortest-path} are called metric or Levi-Civita geodesics. 

The connection  $\widetilde{\Gamma}^{\lambda}_{\,\,\,\mu\nu}$ of modified symmetric teleparallel gravity defines the  parallel transport of vectors.
The vector $v^{\mu}(x)$ is said to be parallel transported along a curve $x^{\mu}(\tau)$, whose tangent vector is $u^{\mu}=dx^{\mu}/d\tau$, if the following equation is satisfied: 
\begin{equation}
\label{parallel-transport}
u^{\nu} \widetilde{\nabla}_{\nu} v^{\mu} = 0
\end{equation}
along the curve. A curve is said to be  autoparallel if its tangent vector is parallel
transported along itself, namely $u^{\nu} \widetilde{\nabla}_{\nu} u^{\mu} = 0$. This is equivalent to 
\begin{equation}
\label{geodesics-autoprallel}
\frac{d^2x^{\mu}}{d\tau^2} +   \widetilde{\Gamma}^{\mu}_{\,\,\, \alpha\beta} \frac{dx^{\alpha}}{d\tau} \frac{dx^{\beta}}{d\tau} =0. 
\end{equation}   
The curves that satisfy this equation are called straightest paths or autoparallel geodesics. 

Since in modified symmetric teleparallel gravity metric geodesics are different from  autoparallel geodesics, the question arises as to which kind of geodesics describe the free fall of a test particle.

\section{Equivalence principle and free fall}\label{freefall}

Einstein's principle of equivalence is the heuristic foundation of the general theory of relativity \cite{Einstein,Pauli,Norton}. Although it was originally formulated in 1907  by Einstein \cite{Einstein}  only for a constant gravitational field, its infinitesimal version for arbitrary gravitational fields seems to have been first enunciated in 1921 by Pauli \cite{Pauli}. There are at least three different formulations of the equivalence principle with increasing degree of stringency \cite{Will1,Will2}. Even in its strongest form, the principle has so far passed severe experimental tests \cite{Will1,Will2,Voisin}. Here we shall be concerned with the Einstein equivalence principle as set forth by Weinberg \cite{Weinberg}.

Let  $\eta_{\mu\nu} =  \mbox{diag} \, (1,-1,-1,-1)$ be the Minkowski metric.  One possible formulation of the   equivalence principle splits it into two parts.

{\bf EP1} At every spacetime point $X$ in an arbitrary gravitational field there exists a {\it locally inertial reference frame } in which ${g_{\mu\nu}}(X)= \eta_{\mu \nu}$ and the connection vanishes at $X$.

{\bf EP2} Within a sufficiently small neighbourhood of the point $X$, in the locally inertial reference frame the laws of nature are those as written in  special relativity in the absence of gravitation.

The first part is a mathematical theorem \cite{Weinberg} that holds true as long as the connection is symmetric, which is the case in $f(Q)$ gravity. The second part is a physical hypothesis that allows one to find the equation of motion of a freely falling test particle in an arbitrary coordinate system.

 Let $x^{\mu}$ be Cartesian coordinates attached to the locally inertial reference frame at point $X$.  According to EP2, the equation of motion of a free test particle --- a particle which is not subject to any non-gravitational force --- is    
\begin{equation}
\label{free-particle-local-inertial}
\frac{d^2x^{\mu}}{d\tau^2} = 0, 
\end{equation} 
where the proper time $d\tau$ is defined by
\begin{equation}
\label{proper-time-inertial}
d\tau^2 = \eta_{\mu\nu} dx^{\mu}dx^{\nu}.
\end{equation} 
Now let ${\hat x}^{\mu}$ be any other coordinate system at $X$, meaning that ${\hat x}^{\mu} = {\hat x}^{\mu}(x)$ and, conversely, $x^{\mu} = x^{\mu}({\hat x})$. Then, the  chain  rule shows that, in terms of the new coordinates, the equation of motion \eqref{free-particle-local-inertial} becomes \cite{Weinberg}
\begin{equation}
\label{free-particle-local-arbitrary-coordinates}
\frac{d^2{\hat x}^{\lambda}}{d\tau^2} + \frac{\partial {\hat x}^{\lambda}}{\partial x^{\sigma}} \frac{\partial^2 x^{\sigma}}{\partial {\hat x}^{\mu} \partial {\hat x}^{\nu}} \frac{d{\hat x}^{\mu}}{d\tau}\frac{d{\hat x}^{\mu}}{d\tau} = 0.
\end{equation} 
Since the connection vanishes at $X$ in the locally inertial coordinate system $x^{\mu}$, the connection transformation law \eqref{connection-transformation} shows that the coefficient of the quadratic term in the velocities in Eq. \eqref{free-particle-local-arbitrary-coordinates} is the connection in the arbitrary coordinate system ${\hat x}^{\mu}$. But, because all connections transform the same way, without additional assumptions one cannot guarantee that the connection that appears in \eqref{free-particle-local-arbitrary-coordinates} is the Levi-Civita connection. The present reasoning is unable  to decide whether a freely falling test particle follows a metric geodesic or an autoparallel geodesic.

With the additional hypothesis that the locally inertial coordinate system can be so chosen that the first derivatives of $g_{\mu\nu}(x)$ vanish at $X$, it is proved that the connection that appears in Eq. \eqref{free-particle-local-arbitrary-coordinates} is necessarily the Levi-Civita connection \cite{Weinberg}. Accordingly,  Weinberg enlarged the equivalence principle by making one more assumption \cite{Weinberg}. 

{\bf EP3} The locally inertial reference frame at $X$ can be chosen in such a way that the first derivatives of the metric tensor vanish at $X$.

The thus extended principle of equivalence entails that the Levi-Civita connection is selected.   
Note, however,  that in modified symmetric teleparallel gravity Eq. \eqref{Nonmetricity} implies that the connection and the first derivatives of the metric cannot vanish at the same time at every point $X$ because the nonmetricity tensor is nonzero.  Thus, since EP3 fails to be true, Weinberg's proof is not applicable in this modified theory of gravity and the Levi-Civita connection is not necessarily singled out. We have to look somewhere else in order to decide between metric and autoparallel geodesics.

\section{Bianchi identity and free fall}\label{Bianchi-identity}

Let us denote by  ${\cal M}_{\mu \nu}$  the left-hand side of Eq \eqref{f(Q)-equations} and define  ${\cal C}_{\alpha}$ as the left-hand side of Eq. \eqref{connection-equations}:
\begin{eqnarray}
\label{left-hand-side-metric}
{\cal M}_{\mu \nu} & \overset{\mbox{\tiny def}}{=} &  \frac{2}{\sqrt{-g}}\widetilde{\nabla}_\alpha\bigl(\sqrt{-g}f_QP^{\alpha}_{\,\,\,\,\mu \nu}\bigr)- \frac{1}{2}fg_{\mu \nu}+f_Q\bigl(P_{(\mu \vert\alpha \beta}Q_{\nu )}^{\,\,\,\alpha \beta}-2P_{\alpha \beta (\mu}Q^{\alpha \beta}_{\,\,\,\,\,\,\,\, \nu )}\bigr), \\
\label{left-hand-side-connection}
{\cal C}_{\alpha} & \overset{\mbox{\tiny def}}{=} &  \widetilde{\nabla}_{\mu}\widetilde{\nabla}_{\nu} \bigl( \sqrt{-g}f_Q P^{\mu\nu}_{\quad \alpha}\bigr).
\end{eqnarray}

The generalized Bianchi identity
\begin{equation}
\label{Bianchi}
\nabla_{\mu}{\cal M}^{\mu}_{\,\,\,\, \nu}  +  {\cal C}_{\nu} = 0 
\end{equation} 
holds \cite{Dambrosio,Lavinia-review} as a consequence of the diffeomorphism invariance of the action \eqref{action}. It is a crucial fact that the covariant derivative that enters the above Bianchi identity is the one defined by the Levi-Civita connection.

Combined with the field equations \eqref{f(Q)-equations} and \eqref{connection-equations}, the Bianchi identity \eqref{Bianchi} implies that the energy momentum tensor satisfies 
\begin{equation}
\label{energy-momentum-tensor-conservation}
\nabla_{\mu}T^{\mu}_{\,\,\,\, \nu}  = 0.
\end{equation} 
Let us assume that the energy-momentum tensor describes a structureless test particle, meaning that $T^{\mu \nu}$ is concentrated on the particle's worldline and the monopole moment $\int T^{\mu \nu} d^3x \neq 0$ whereas the dipole moment and all higher multipole moments of $T^{\mu \nu}$ vanish. Then it is proved \cite{Papapetrou} that  Eq. \eqref{energy-momentum-tensor-conservation} entails that a freely falling test particle follows a metric geodesic described by Eq. \eqref{geodesics-shortest-path}. Papapetrou's proof is not particularly difficult but it can be made quite elementary  by letting the energy momentum tensor be $T^{\mu \nu} = \rho_0(x)  u^{\mu}u^{\nu}$, which describes  a fluid of noninteracting particles with proper energy density $\rho_0$ and four-velocity $u^{\mu}$. This is tantamount to  treating each test particle as  a speck in a cloud of dust \cite{Adler}.

\section{Conclusion}\label{Conclusion}

Our initial question has been answered: in modified symmetric teleparallel gravity a freely falling test particle follows a metric geodesic, not an autoparallel geodesic, although the theory fails to obey the equivalence principle in the sense of Weinberg. This property of freely falling test particles is also of consequence regarding the extent to which, under suitable conditions, this modified theory of gravity  is equivalent to general relativity. If $f(Q)=AQ+B$ or $Q$ is constant then modified symmetric teleparallel gravity is equivalent to general relativity not only at the level of the field equations but also at the level of the equations of motion of freely falling test particles. Under the above conditions the two theories are truly physically equivalent.

Last but not least, it is worth stressing that modified symmetric teleparallel gravity establishes that Weinberg's hypothesis EP3 is a sufficient but not necessary condition for singling out the Levi-Civita connection.

\begin{acknowledgments}

The author is thankful to Luca Moriconi of Universidade Federal do Rio de Janeiro (UFRJ) who, during a seminar, asked  the question that motivated this unpretentious study. 

\end{acknowledgments}

\end{document}